\documentstyle[pra,aps,epsfig,array,twocolumn]{revtex}
\begin{document}
\def\eps{\epsilon}
\newcommand{\ket}[1]{| \, #1 \rangle}
\newcommand{\bra}[1]{ \langle #1 \,  |}
\newcommand{\proj}[1]{\ket{#1}\bra{#1}}
\newcommand{\braket}[2]{\langle #1 | #2 \rangle}
\newcommand{\pket}[2]{ | \, #1 \rangle \otimes | \, #2  \rangle  }
\newcommand{\pbra}[2]{ \langle #1\,  | \otimes \langle #2 \,  |}
\newcommand{\ppket}[3]{ | \, #1 \rangle \otimes  | \, #2 \rangle
                        \otimes  | \, #3 \rangle }
\newcommand{\pproj}[2]{\ket{#1}\bra{#1}\otimes\ket{#2}\bra{#2} }
\newcommand{\abs}[1]{ | \, #1 \,  |}
\newcommand{\av}[1]{\langle\,#1\,\rangle}
\newcommand{\asred}{\preceq}
\newcommand{\asredchem}{\leadsto}
\newcommand{\asequ}{\approx}
\newcommand{\exred}{\le}
\newcommand{\exredchem}{\to}
\newcommand{\exequ}{\equiv}
\newcommand{\exequchem}{\rightleftharpoons}
\newcommand{\stored}{\lesssim}
\newcommand{\tr}[2]{{\rm tr}_{\rm \scriptscriptstyle #1}(#2)}
\newcommand{\identity}{\mbox{\boldmath{1} \hspace{-0.33cm} \boldmath{1} }}
\newcommand{\PT}[2]{(#1)^{\rm T_{#2}}}
\newcommand{\upp}[1]{^{\rm \scriptscriptstyle #1}}
\newcommand{\dnn}[1]{_{\rm \scriptscriptstyle #1}}
\newcommand\bea{\begin{eqnarray}}
\newcommand\eea{\end{eqnarray}}
\newcommand{\beq}{\begin{equation}}
\newcommand{\eeq}{\end{equation}}
\newtheorem{lem}{Lemma}
\newtheorem{theo}{Theorem}
\newtheorem{dfn}{Definition}
\newtheorem{cor}{Corollary}

\twocolumn[\hsize\textwidth\columnwidth\hsize\csname
@twocolumnfalse\endcsname
\title{Notes on Landauer's Principle, Reversible Computation,
and Maxwell's Demon}
\author{Charles H. Bennett}
\address{IBM Research Division, Yorktown Heights, NY 10598, USA ---
{\tt bennetc@watson.ibm.com}}

\date{\today}
\maketitle
\begin{abstract}Landauer's principle, often regarded as the basic
principle of the thermodynamics of information processing, holds
that any logically irreversible manipulation of information, such
as the erasure of a bit or the merging of two computation paths,
must be accompanied by a corresponding entropy increase in non-
information-bearing degrees of freedom of the information
processing apparatus or its environment. Conversely, it is
generally accepted that any logically reversible transformation of
information can in principle be accomplished by an appropriate
physical mechanism operating in a thermodynamically reversible
fashion.  These notions have sometimes been criticized either as
being false, or as being trivial and obvious, and therefore
unhelpful for purposes such as explaining why Maxwell's Demon
cannot violate the Second Law of thermodynamics.  Here I attempt
to refute some of the arguments against Landauer's principle,
while arguing that although in a sense it is indeed a
straightforward consequence or restatement of the Second Law, it
still has considerable pedagogic and explanatory power, especially
in the context of other influential ideas in 19'th and 20'th
century physics. Similar arguments have been given by Jeffrey
Bub~\cite{Bub02}
\end{abstract}\bigskip\bigskip
]

\vspace{1cm}
\subsection*{Landauer's Principle}
In his classic 1961 paper~\cite{Landauer61}, Rolf Landauer
attempted to apply thermodynamic reasoning to digital computers.
Paralleling the fruitful distinction in statistical physics
between macroscopic and microscopic degrees of freedom, he noted
that some of a computer's degrees of freedom are used to encode
the logical state of the computation, and these "information
bearing" degrees of freedom (IBDF) are by design sufficiently
robust that, within limits, the computer's logical (i.e. digital)
state evolves deterministically as a function of its initial
value, regardless of small fluctuations or variations in the
environment or in the computer's other non-information-bearing
degrees of freedom (NIBDF).  While a computer as a whole
(including its power supply and other parts of its environment),
may be viewed as a closed system obeying reversible laws of motion
(Hamiltonian or, more properly for a quantum system, unitary
dynamics), Landauer noted that the logical state often evolves
irreversibly, with two or more distinct logical states having a
single logical successor. Therefore, because Hamiltonian/unitary
dynamics conserves (fine-grained) entropy, the entropy decrease of
the IBDF during a logically irreversible operation must be
compensated by an equal or greater entropy increase in the NIBDF
and environment. This is Landauer's principle. Typically the
entropy increase takes the form of energy imported into the
computer, converted to heat, and dissipated into the environment,
but it need not, since entropy can be exported in other ways, for
example by randomizing configurational degrees of freedom in the
environment.

Landauer's principle appears straightforward, but there is some
subtlety in understanding when it leads to thermodynamic
irreversibility. If a logically irreversible operation like
erasure is applied to random data, the operation still may be
thermodynamically reversible, because it represents a reversible
transfer of entropy from the data to the environment, rather like
the reversible transfer of entropy to the environment when a gas
is compressed isothermally. But if, as is more usual in computing,
the logically irreversible operation is applied to known data, the
operation is thermodynamically irreversible, because the
environmental entropy increase not compensated by any decrease of
entropy of the data.  This wasteful situation, in which an
operation that {\em could\/} have reduced the data's entropy is
applied to data whose entropy is already zero, is analogous to the
irreversibility that occurs when a gas is allowed to expand
freely, without doing any work, then isothermally compressed back
to its original volume.  Fortunately, these wasteful operations
can be entirely avoided: it is possible to reprogram any
deterministic computation as a sequence of logically reversible
steps, provided the computation is allowed to save a copy of its
input.  The logically reversible version of the computation, which
need not use much more time or memory than the original
irreversible computation it simulates, can then, at least in
principle, be performed in a thermodynamically reversible fashion
on appropriate hardware.

\subsection*{Objections to Landauer's Principle}

One of the main objections to Landauer's principle, and in my
opinion the one of greatest merit, is that raised by Earman and
Norton~\cite{Exorcist1ff}, who argue that since it is not
independent of the Second Law, it is either unnecessary or
insufficient as an exorcism of Maxwell's demon.  I will discuss
this objection further in the third section.

Others have argued that Landauer's principle is actually false or
meaningless.  These objections are of three kinds:
\begin{enumerate} \item It is false or meaningless because there
is no connection between thermodynamic quantities like heat and
work and mathematical properties like logical reversibility, so
that comparing the two is comparing apples and oranges;
\item It (or more precisely its converse) is false because {\em all\/}
data processing operations, whether logically reversible or not,
require the dissipation of at least $kT\ln 2$ of energy---and
indeed usually much more---to be accomplished by any actual
physical apparatus; or \item It is false because even logically
irreversible operations can in principle be accomplished without
an entropy increase in the environment.
\end{enumerate}

The first objection touches deep questions of the relation between
mind and matter which are not entirely in the province of science,
although physicists have long felt a need to address them to some
extent.  From its beginning, the history of the Maxwell's Demon
problem has involved discussions of the role of the Demon's
intelligence, and indeed of how and whether one ought to
characterize an "intelligent being" physically.  On this question
I will take the usual approach of physicists, and banish questions
about intelligence by substituting an automatically functioning
mechanism whenever an intelligent being is required. Not only is
this mechanism supposed to be automatic, it ought to obey accepted
physical laws.  In particular, the entire universe, including the
Demon, should obey Hamiltonian or unitary dynamics, when regarded
as a closed autonomous system.  From this viewpoint the first
objection loses much of its persuasiveness, since there appears to
be no deep conceptual problem in inquiring whether an
automatically functioning apparatus designed to process
information, ie a computer, can function in a thermodynamically
reversible fashion, and if not, how the thermodynamic cost of
operating the apparatus depends on the mathematical properties of
the computation it is doing.

The second objection, that even logically reversible
data-processing operations cannot be accomplished in a
thermodynamically reversible fashion, I believe has largely been
overcome by explicit models, proposed by myself and others, of
physical mechanisms, which obey the accepted conventions of
thermodynamic or mechanical thought experiments, and which
accomplish reversible computation at zero cost (so-called
ballistic computers, such as the Fredkin-Toffoli hard sphere
model~\cite{FT82}), or at a per-step cost tending to zero in the
limit of slow operation (so-called Brownian computers, discussed
at length in the review article~\cite{B82}). These questions were
revisited and vigorously debated in an exchange in Phys. Rev.
Lett.~\cite{Porod}, to which the reader is referred. Of course in
practice almost all data processing is done on macroscopic
apparatus,  dissipating macroscopic amounts of energy far in
excess of what would be required by Landauer's principle.
Nevertheless, some stages of biomolecular information processing,
such as transcription of DNA to RNA, appear to be accomplished by
chemical reactions that are reversible not only in principle but
in practice.

Among the most important logically reversible operations are
copying unknown data onto a blank (ie initially zero) register,
and the reverse process, namely erasure of one of two copies of
data known to be identical.  To show that these operations are
logically reversible, we write them in standard programming
notation, where ${\tt y}$ represents the register to be changed
and ${\tt x}$ a reference register which is either added to it or
subtracted from it:

\begin{verbatim}
y := y + x
y := y - x
\end{verbatim}
\medskip\noindent
The first operation copies the value of ${\tt x}$ into ${\tt y}$,
if ${\tt y}$ is initially zero; the second erases ${\tt y}$ to
zero, if ${\tt y}$ and ${\tt x}$ are known to be initially equal.
For any initial values of ${\tt x}$ and ${\tt y}$ the two
operations are logically reversible, since each exactly undoes the
effect of the other.  Physically reversible means of performing
these and other logically reversible operations are discussed in
\cite{Landauer61},\cite{B82}.  An important example of a logically
reversible operation, in the context of Maxwell's demon and
Szilard's engine (cf.~Fig.~12 in \cite{B82}), is reversible
measurement, in particular the reversible transition of a memory
element from a standard initial state S into one of two states,
call them L and R, depending on whether the single molecule in
Szilard's engine is located on the left or right.  The physical
reversibility of this operation in terms of phase space volumes is
illustrated in Fig.~12 of \cite{B82}; an explicit clockwork
mechanism for carrying out this reversible measurement is analyzed
in \cite{B87}.

For the remainder of this section I will focus on the third
objection, the argument that even logically irreversible
operations can be accomplished without an entropy increase in the
environment. This position has been asserted in various forms by
Earman and Norton~\cite{Exorcist16ff} and by
Shenker~\cite{Shenker}, not as an argument against the second law,
but rather as an argument against the thesis of reversible
measurement.

In the context of a modified Szilard engine, similar to figure 12
of \cite{B82}, Earman and Norton consider a demon with memory
capable of two states, L and R, which follows a program of steps
similar to a computer program, executing one of two separate
logically reversible subprograms, ``program-L'' and ``program-R'',
depending on whether the molecule is found on the left or right
side of the partition. Initially, they assume the memory is in
state L (their state L does double duty as the standard initial
memory state, called S in \cite{B82} and as the left
post-measurement state, called L in \cite{B82}).  Instructions are
labelled below according to the part of the demon's cycle. Steps
M1-M4 comprise the measurement segment, R1-R5 the ``right"
subroutine, and L1-L4  the ``left" subroutine.  The memory state
at the end of the each operation is shown in brackets.
\begin{verbatim}
 M1. Insert partition [L]
 M2. Observe the particle's chamber[L] or [R]
 M3. If memory bit =R, go to R1 [R]
 M4. If memory bit = L, go to L1 [L]
 R1. Attach pulleys so right chamber
     can expand [R]
 R2. Expand,doing isothermal work W [R]
 R3. Remove pulleys [R]
 R4. Transform known memory bit from R to L [L]
 R5. Go to M1 [L]
 L1. Attach pulleys so left chamber
     can expand [L]
 L2. Expand,doing isothermal work W [L]
 L3. Remove pulleys [L]
 L4. Go to M1 [L]
\end{verbatim}

According to the thesis of reversible measurement, which Earman
and Norton dispute, step M2 can be accomplished reversibly,
without any entropy increase in the environment. Temporarily
granting this assumption, Earman and Norton argue that at the
culmination of step R4 or L3, both the gas and demon are back in
their initial states. Work W has been done by removing energy heat
from the reservoir without any other change in the universe. If
all this were really so the second law would have been violated.

I would argue on the contrary that while each of the routines L
and R by itself is logically reversible, the combination is not
logically reversible, because it includes a merging in the flow of
control, which is just as much a case of logical irreversibility
as the explicit erasure of data. The instruction M1 has two
predecessors (L4 and R5). Therefore, when executing this program,
there is a two to one mapping of the logical state as control
passes from L4 or R5 to M1.  This is where the work extracted by
the demon must be paid back, according to Landauer's principle.
There is then no need to seek an additional entropy increase
elsewhere in the cycle, such as in the measurement step M2, whose
reversibility has in any case been demonstrated for explicit
models~(\cite{B82}\cite{B87}).

The fact that a merging of the flow of control constitutes logical
irreversibility is also illustrated in fig.~1 of ref.~\cite{B82},
where the final Turing machine configuration has two predecessors
not because of an explicit erasure of data, but because the head
could have arrived at its final location by two different shifts,
one of which corresponds to the intended computation path, while
the other corresponds to an allowed transition to the final state
from an extraneous predecessor configuration which is not part of
the intended computation.

A similar objection to Landauer's principle, this time illustrated
with an explicit gear mechanism involving a pinion operating
between two rigidly connected but opposing half-racks, is
presented by Shenker~\cite{Shenker} in her figure 5, which is
adapted from a diagram of Popper (see \cite{LeffRex}, chapter 1,
Fig. 12). The accompanying text argues that a single external
manipulation, namely a counterclockwise rotation applied to pinion
gear B, would restore a memory element called the "key" from
either of two initial positions labelled respectively "R" and "L",
to a neutral final position labelled "?", corresponding to the
standard state "S" in \cite{B87}.

While it is indeed true that the counterclockwise rotation of the
pinion would do this, the act of performing that rotation is not
thermodynamically reversible, as one can see by considering in
more detail the possible motions of and mutual constraints between
the two relevant mechanical degrees of freedom, namely 1) the
rotation angle of the pinion and 2) the lateral (left-right)
displacement of the key with its two rigidly attached half-racks,
termed ``grooves for gear B" in the figure.  For any rotation
angle of the pinion, there will be a finite range of backlash
within which the key can rattle back and forth before being
stopped by collision with the gear teeth on the pinion (to demand
zero backlash would entail an infinite negative configurational
entropy; the argument given here in support of Landauer's
principle is independent of the amount of backlash, requiring only
that it be nonzero).  Consider the resetting action when the key
is initially in the ``L" position.  The accompanying figure shows
schematically how the range of motion of the information bearing
coordinate (in this case the left/right displacement of the key)
varies as a function of a controlling coordinate (in this case the
rotation angle of the pinion B) whose steady increase brings about
a merger of two logically distinct paths of the information
processing apparatus (in this case the resetting of the key to the
neutral position).  At stage 1, the information bearing

\vbox{
\begin{figure}
\epsfxsize=6.8cm \epsfbox{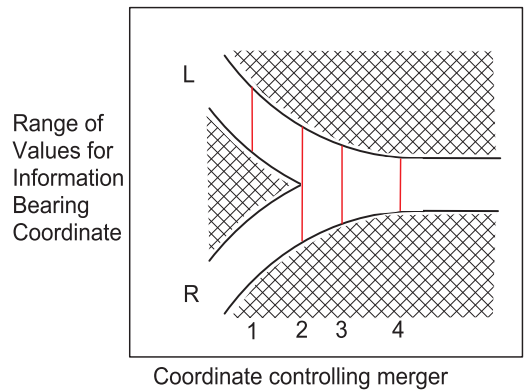}
\smallskip
\caption{Merging of two computation paths.} \label{merge2}
\end{figure}
} \noindent coordinate is confined to one of the two
about-to-be-merged paths.  At stage 2, the barrier separating
these paths disappears, and the information bearing coordinate
suddenly gains access to twice as large a range as it had before.
This is an irreversible entropy increase of $k \ln 2$ analogous to
the $Nk\ln 2$ entropy increase when an ideal gas of N atoms leaks
by free expansion into twice its original volume, without doing
any work. At stage 3, the controlling coordinate (pinion) does
work on the information bearing coordinate (key), eventually, by
stage 4, compressing it back to its original range of motion.

Although Landauer's principle appears simple to the point of
triviality, there is, as noted in the first section, a somewhat
subtle distinction in how it applies depending on the nature of
the initial state. As noted in \cite{B82}, fig. 16, a logically
irreversible operation, such as the erasure of a bit or the
merging of two paths, may be thermodynamically reversible or not
depending on the data to which it is applied. If it is applied to
random data---a bit that is initially equiprobably distributed
between 0 and 1, or a key that is equiprobably on the R or the L
path in the accompanying figure, it is thermodynamically
reversible, because it increases decreases the entropy of the data
while increasing the entropy of the environment by the same
amount.  In terms of usual thermodynamic thought experiments, it
is analogous to isothermal compression, which decreases the
entropy of a gas while increasing the entropy of the environment.
This situation arises in the application of Landauer's principle
to Szilard's engine: the data being erased (R or L) is random;
therefore its erasure represents a reversible entropy transfer to
the environment, compensating an earlier entropy transfer {\it
from\/} the environment during the isothermal expansion phase, and
making the total work yield of the cycle zero, in obedience to the
Second Law.   However, as Landauer and Shenker note, the data in
the course of the usual deterministic digital computation is not
random, but on the contrary determined by the computer's initial
state. Thus, at least in the context of knowledge of the initial
state, whenever a logically irreversible step occurs in the course
of a deterministic computation, one of the predecessors is certain
to have been on the computation path, while the other(s) have zero
chance of having been on the path. Shenker argues (her figure 3)
that the 1:1 sequence of states actually visited in the course of
a deterministic computation means that deterministic computers are
not bound by Landauer's principle.  In fact, by the above
argument, the performance of a 1:1 state mapping by a manipulation
that {\em could have\/} performed a 2:1 mapping is
thermodynamically irreversible, the irreversibility being
associated with the wasteful instant (stage 2 in the above figure)
at which a constrained degree of freedom is allowed to escape from
its constraint without any work being exacted in exchange. In
fact, the significance of Shenker's and Landauer's observation
(viz.~that the states actually visited in a deterministic
computation comprise an unbranched chain) is somewhat different.
It means that it is always possible to globally reprogram any
computation that saves a copy of its input as a sequence of
logically reversible steps, and therefore to perform it in a
thermodynamically reversible fashion, even though the original
computation, before this reprogramming, would not have been
thermodynamically reversible.  The earliest example of reversible
programming by Lecerf~\cite{Lecerf} did not attract much
attention, perhaps because it treated reversibility as a purely
mathematical property without considering its physical
implementation or significance. Physical models which can execute
reversibly programmed computations in a thermodynamically
reversible fashion are discussed in ~\cite{B73}, \cite{FT82} and
\cite{B82}. Efficient reversible programming techniques, and the
space/time tradeoff for reversible simulation of irreversible
computations, have received much study~\cite{Revcomp}.

Another critic of Landauer's principle, Thomas
Schneider~\cite{Schneider}, argues that the energy cost of
biological information processing ought not to be regarded as
coming from resetting or erasure, but rather from a two step
process in which a molecular computing system is first ``primed''
or activated by the addition of energy, and later dissipates this
energy by falling into one of several final states.  I would argue
on the one hand this is not inconsistent with Landauer's
principle, and on the other hand that not all biochemical
information processing systems are best viewed in this way; for
example the transcription from DNA to RNA, after strand
initiation, can be viewed as a logically reversible copying
process driven chiefly not by prior activation of the reactant,
but by removal of one of the reaction products, pyrophosphate.

Landauer's principle applies in different ways to the several
kinds of physical system capable of reversible computation or
molecular scale thermodynamic engines. Such devices are broadly of
three types:
\begin{itemize}
\item Ballistic computers: conservative dynamical systems like
Fredkin's billiard ball computer, which follows a mechanical
trajectory isomorphic to the desired computation. Such systems are
incapable of merging of trajectories, so they can only be
programmed to do logically reversible computations, and indeed do
so at constant velocity without dissipating any energy.  These
devices must be isolated from external heat baths or noise
sources, so they are not directly relevant to the Maxwell's demon
problem.
\item Externally clocked Brownian machines, in which a
control parameter is quasistatically varied by a macroscopic
external agency to drive the system through its sequence or cycle
of operations. All other coordinates are free to move randomly and
equilibrate themselves within the constraint set by the value of
the control parameter, and are typically coupled to a thermal
bath. Many realizations of Szilard's engine, including those
discussed by Earman and Norton, and that depicted in Fig. 1 of the
present paper, are of this type, as are most of the proposed
realizations of quantum computers. The thermodynamic cost of
operating such a machine is the work done by the external agency,
integrated over the cycle or sequence of operations. Mergings of
trajectories are possible, but are thermodynamically costly if
applied to nonrandom data, by the arguments given above.  If such
mergings are avoided through reversible programming, these devices
operate reversibly in the usual sense of thermodynamic thought
experiments, ie their dissipation per step is proportional to the
driving force, tending to zero in the limit of zero speed of
operation.

\item Fully Brownian machines, in which all coordinates are allowed
to drift freely, within the constraints they mutually exert on one
other.  This kind of machine is like an externally clocked
Brownian machine where the external agency has ``let go of the
handle'', and attached a weak spring, so the control coordinate
can drift randomly forward or backward. Examples include Gabor's
engine~\cite{Gabor}, Feynman's classic ratchet and
pawl~\cite{Feynman}, Bennett's enzymatic and clockwork computers
(\cite{B82}, figs.~6-9), and enzymes such as RNA polymerase and
polynucleotide phosphorylase. The thermodynamic cost of operating
this type of machine is determined by the weakest spring
sufficient to achieve a net positive drift velocity.  Because this
type of device can drift backward and explore the tree of logical
predecessors of states on its intended computation path, a small
amount of merging is less costly than it would be for an
externally clocked Brownian device; as described in more detail in
(\cite{B82}, fig.~10) and at the end of \cite{B73}, such a device
will drift forward, although perhaps very slowly, if driven
strongly enough to prevent infinite backward excursions.
\end{itemize}

Most discussions of the thermodynamics of information processing,
particularly in Brownian computers, ignore hardware errors, i.e.
transitions to states that are neither on the intended computation
path nor predecessors of states on the intended path. Such
transitions, which in a genetic system would correspond to
mutations or copying errors, occur in any natural computing
apparatus but are ruled out by the infinite potential energy
barriers in idealized models such as the clockwork computer of
~\cite{B82}. The theory of fault-tolerant computation, which has
recently received a great deal of study in the context of quantum
computing, aims to show how, through appropriate design and
software, arbitrarily large reliable computations can be performed
on imperfect hardware suffering some fixed rate of hardware
errors. A full understanding of the thermodynamics of information
processing ought therefore to include an understanding of the
thermodynamics of error correction in a fault tolerant setting, a
subject that appears to be in its infancy. A tractable toy example
of fault tolerance, namely a Brownian copying system patterned on
the proofreading enzymes involved in DNA replication, was studied
in \cite{B79}, and its dissipation/error tradeoff analyzed. A
related problem arises with liquid state NMR quantum computation:
although one can make the gates fairly reliable, only reversible
gates are available and the initial thermal state is almost
perfectly random. Schulman and Vazirani \cite{SV98} have devised
an algorithm for pumping the nonrandomness in a large number of
thermal qubits into a few of them, leaving the rest perfectly
random.  Much work remains to be done on the dissipation/error
tradeoff in more general settings.

\subsection*{Landauer's principle in the context of other
ideas in 19'th and 20'th century physics}  Earman and Norton have
pointed out with some justice that Landauer's principle appears
both unnecessary and insufficient as an exorcism Maxwell's demon,
because if the Demon is a thermodynamic system already governed by
the Second Law, no further supposition about information and
entropy is needed to save the Second Law.  On the other hand, if
the Demon is not assumed to obey the Second Law, no supposition
about the entropy cost of information processing can save the
Second Law from the Demon.

I would nevertheless argue that Landauer's principle serves an
important pedagogic purpose of helping students avoid a
misconception that many people have fallen into during the 20'th
century, including giants like von Neumann, Gabor, and Brillouin
and even, perhaps, Szilard\footnote{Szilard's classic 1929 paper
~\cite{Szilard} is tantalizingly ambiguous in this respect.  While
most of the paper seems to attribute an irreducible thermodynamic
cost to information acquisition---and this indeed is how his paper
has usually been subsequently interpreted---the detailed
mathematical analysis at the end (paragraphs preceding his eq.~21)
shows the entropy increase as occurring during the resetting step,
in accordance with Landauer's principle. Probably Szilard thought
it less important to associate the entropy increase with a
particular stage of the cycle than to show that it must occur
somewhere during the cycle.}.  This is the informal belief that
there is an intrinsic cost of order $kT$ for every elementary act
of information processing, e.g. the acquisition of information by
measurement, or the copying of information from one storage medium
into another, or the execution of a logic operation by a computer,
regardless of the act's logical reversibility or irreversibility.
In particular, the great success of the quantum theory of
radiation in the early 20'th century led Gabor and Brillouin to
seek an exorcism of the Demon based on a presumed cost of
information acquisition, which in turn they attributed to the
energy cost of a thermal photon, or in the case of Gabor's
high-compression Szilard engine~\cite{Gabor}, to the cost of
recreating a static radiation field localized to one end of a long
cylinder, into which the molecule would wander to trigger the
power stroke.  Landauer's principle, while perhaps obvious in
retrospect, makes it clear that information processing and
acquisition have no intrinsic, irreducible thermodynamic cost,
whereas the seemingly humble act of information destruction does
have a cost, exactly sufficient to save the Second Law from the
Demon.  Thus measurement and copying {\it can\/} be intrinsically
irreversible, but only when they are conducted in such a way as to
overwrite previous information.

The Second Law, uniquely among physical principles, is and
probably always will be in need of explanations and worked-out
examples showing why microscopically reversible physical systems
cannot escape it.  When first told of the Demon, any normally
curious person will be dissatisfied with the explanation ``It
can't work because that would violate the Second Law'' and will
want to see exactly {\em why\/} it can't work.  That is the virtue
of worked-out examples such as Feynman's ratchet and pawl. Indeed
Feynman's ratchet and pawl argument provides a more fundamental
and elegant refutation of Gabor's engine than that given by Gabor,
one that does not depend on the quantum theory of radiation.
Gabor's engine (\cite{Gabor}, his fig. 7) uses an optically
triggered mechanism to trap the molecule at one end of a long
cylinder. Once the molecule has been trapped, it is made to do a
large amount, $k_BT\ln X$ of work by isothermal expansion, where
$X$ is the expansion ratio, hopefully more than enough to replace
the energy $E$ dissipated when the trap was sprung. But no
trapping mechanism, optical or otherwise, can be completely
irreversible. As Feynman points out in his analysis of the fall of
the pawl off the edge of a ratchet tooth, which is designed to
prevent the ratchet from rotating backwards, a trapping mechanism
that dissipates energy $E$ at temperature $T$ has a probability
$\exp(-E/k_BT)$ of running backward.  The cyclic operation of
Gabor's engine is mathematically equivalent to a ratchet machine
which dissipates the trapping energy $E$ when the pawl falls, and
does $k_BT\ln X$ work as the ratchet rotates in the intended
forward direction to the next tooth.  If $E>k_BT\ln X$, Gabor's
engine will run in the intended forward direction, but will not do
enough work to replace the energy lost in springing the trap. If
$E<k_BT\ln X$, the engine will run backwards of its intended
direction, alternately compressing the molecule and letting it
escape into the long cylinder through backward operation of the
trap, and again it will not violate the second law.

\section*{Acknowledgements}
I acknowledge support from the US Army Research office, grant
DAAG55-98-C-0041 and DAAG55-98-1-0366.


\begin{references}
\bibitem{Bub02} Jeffrey Bub, ``Maxwell's Demon and the Thermodynamics
of Computation'' (2002), arXiv:quant-ph/0203017 .
\bibitem{Landauer61} R. Landauer, ``Dissipation and Heat
Generation in the Computing Process'' {\em IBM J. Research and
Develop.\/} {\bf 5,} 183-191 (1961), reprinted in \cite{LeffRex},
pp. 188-196 (1st edition), 148-156 (2nd edition).
\bibitem{Exorcist1ff} J. Earman and J.D. Norton, ``Exorcist XIV: The
Wrath of Maxwell's Demon. Part II. From Szilard to Landauer and
Beyond,'' Studies in the History and Philosophy of Modern Physics
30, 1-40 (1999), cf. pages 1-8.
\bibitem{FT82}E. Fredkin and T. Toffoli, `Conservative Logic,''
{\em International Journal of Theoretical Physics\/} 21, 219
(1982), eprint available at {\tt http://digitalphilosophy.org
/download\_documents/ConservativeLogic.pdf }
\bibitem{B82}C.H.Bennett,``The Thermodynamics of Computation---a
Review,''International Journal of Theoretical Physics 21, 905-940
(1982). Reprinted in \cite{LeffRex} pp. 213-248 (1st ed.), 283-318
(2nd ed.
);
{\tt http://www.research.ibm.com \\
/people/b/bennetc/bennettc1982666c3d53.pdf}
\bibitem{Porod} Porod, W., Grondin, R. O., Ferry, D. K. \& Porod, G.
``Dissipation in computation'', {\em Phys.Rev.Lett.,\/} {\bf 52,}
232-235, (1984) and ensuing discussion.
\bibitem{B87} C.H. Bennett, ``Demons, Engines, and the Second
Law,'', {\em Sci. Amer.\/} {\bf 257} 108-117 (1987).
\bibitem{Exorcist16ff} Earman and Norton {\em op. cit.\/} pp. 16-18.
\bibitem{Shenker} Orly R. Shenker, ``Logic and Entropy'',
Philosophy of Science Archive~(2000), eprint at \\
{\tt http://philsci-archive.pitt.edu/documents
/disk0/00/00/01/15/index.html}
\bibitem{LeffRex} Harvey S. Leff and Andrew F.
Rex, {\em Maxwell's Demon: Entropy, Information, Computing\/}
(Princeton: Princeton University Press, 1990); Second edition:{\em
Maxwell's Demon 2: Entropy, Classical and Quantum Information,
Computing\/} (Institute of Physics Publishing, 2003.
\bibitem{Lecerf}Yves Lecerf, Machines de Turing r\'{e}versibles.
R\'{e}cursive insolubilit\'{e} en $n\in N$ de l'equation
$u=\theta^n u$ o\`{u} $\theta$ est un ``isomorphism de codes".
Comptes Rendus Hebdomedaires des S\'{e}ances de L'Acad\'{e}mie
Francaise des Sciences, {\bf 257,} 2597-2600 (1963).
\bibitem{B73} C.H. Bennett, ``Logical Reversibility of Computation'',
{\em IBM J. Research and Develop.\/} {\bf 17} 525-532 (1973),
reprinted in \cite{LeffRex}, pp. 197-204 (1st edition).
\bibitem{Revcomp}
C.H. Bennett ``Time/Space Trade-offs for Reversible Computation"
SIAM J. Computing 18, 766-776 (1989); R.Y. Levine and A.T.
Sherman, ``A note on Bennett's time-space trade-off for reversible
computation. SIAM J. Comput 19, 673-677 (1990); K.J. Lange, P.
McKenzie, and A. Tapp, ``Reversible space equals deterministic
space", {\em J. Comput. System Sci.\/} {\bf 60:2,} 354-367 (2000);
H. Buhrman, J. Tromp, and P. Vitanyi, ``Time and Space Bounds for
Reversible Simulation'' eprint {\tt
www.cwi.nl/\~{}paulv/papers/tradeoff.ps}; R. Williams,
``Space-Efficient Reversible Simulations'' DIMACS REU report
eprint {\tt http://dimacs.rutgers.edu/\~{}ryanw/} (2000)
\bibitem{Schneider} T.D. Schneider ``Sequence Logos, Machine/Channel
Capacity, Maxwell's Demon, and Molecular Computing: a Review of
the Theory of Molecular Machines'' {\em Nanotechnology\/} {\bf 5,}
1-18 (1994).
\bibitem{Szilard} Leo Szilard, {\em Z. Physik\/} {\bf 53,} 840-856
(1929); translation by A. Rapoport and M. Knoller ``On the
decrease of entropy in a thermodynamic system by the intervention
of intelligent beings'' reprinted in \cite{LeffRex} pp. 124-133
(1st edition) and pp. 110-119 (2nd edition).
\bibitem{Gabor} Denis Gabor, ``Light and Information'' {\em
Progress in Optics\/} {\bf 1,} 111-153 (1964), relevant part
reprinted in \cite{LeffRex} pp. 148-159 (1st edition).
\bibitem{Feynman} Richard P. Feynman, ``Ratchet and Pawl,'' chapter 46 in
{\em The Feynman Lectures on Physics, vol. 1\/}, edited by R.P.
Feynman, R.B. Leighton, and M. Sands, Addison-Wesley, Reading,
Mass, (1963).
\bibitem{B79} C.H. Bennett, ``Dissipation-Error Tradeoff
in Proofreading" {\em BioSystems\/} {\bf 11,} 85-91 (1979), eprint
available at
{\tt http://www.research.ibm.com/people/b/bennetc \\
/bennettc1979584d6f5e.pdf}
\bibitem{SV98} Leonard J. Schulman and Umesh Vazirani ``Scalable NMR
Quantum Computation'' eprint quant-ph/9804060 and Proc. 31'st ACM
STOC (Symp. Theory of Computing), 322-329, 1999


\medskip
\end{references}
\end{document}